# A Lax Pair for the Dynamics of DNA Modeled as a Shearable and Extensible Elastic Rod
# II. Discretization of the arc length

Yaoming Shi(a), W.M. McClain(b), and John E. Hearst(a)

(a)Department of Chemistry, University of California, Berkeley, CA 94720-1460
(b)Department of Chemistry, Wayne State University, Detroit, MI 48202

Contact: jehearst@cchem.berkeley.edu

**November 27, 2001**

ABSTRACT

In previous work, the dynamics of the elastic rod was recast in a Lax pair formulation, with fiducial arc length s and time t as continuous independent variables. However, the solution of these equations cannot apply directly to a system where the fiducial arc length s is a discrete variable. In this paper, we show how to discretize the continuous s variable in a way that preserves the integrability of the original system. The t parameter is not discretized, so this algorithm will be especially useful for solutions of the s-discrete and t-continuous elastic rod problem, as may occur in problems where the polymeric structure of the DNA is made explicit.

## I. INTRODUCTION

In paper I[1], we derived a Lax Pair for the geometrically exact Hamiltonian equations for the elastic rod, proving that the problem is integrable in general. In this paper we provide a similar Lax pair reformulation, but with integrable discretization of the fiducial arc length s. The other independent variable, the time t, remains continuous. In a following paper III[2], both s and t will be discretized to permit integrable dynamic solutions.

## II. THE CONFIGURATION SPACE OF THE ELASTIC ROD IN 3D

We treat duplex DNA as a bendable, twistable, extensible, and shearable thin discrete elastic rod. Here and elsewhere in this paper the terms "elastic rod" and "DNA" have the same meaning; so do "the centerline of the rod" and "the axis of DNA".

Let $k$ be an integer that discretizes the continuous variable $s$. Thus, we will express any function $f(s,t)$ as $f(k\Delta s,t)$, or more briefly as $f(k,t)$. At a given time $t$ and at each point $s = k\Delta s$ on the centerline $\mathbf{r}(k,t)$ of the rod, a localized Cartesian coordinate frame (or director frame), $\left\{\hat{\mathbf{d}}_1(k,t), \hat{\mathbf{d}}_2(k,t), \hat{\mathbf{d}}_3(k,t)\right\}$, is affixed with the unit vectors $\hat{\mathbf{d}}_1(k,t)$ and $\hat{\mathbf{d}}_2(k,t)$ in the direction of the principal axes of inertia tensor of the rod cross section. The third unit vector $\hat{\mathbf{d}}_3(k,t)$ is in the normal direction of the cross section. Because the shear is included, the unit vector $\hat{\mathbf{d}}_3(k,t)$ does not necessarily coincide with the tangent vector $\mathbf{\Gamma}(k,t)$ of the centerline of the elastic rod $\mathbf{r}(k,t)$.

Unit vectors $\hat{\mathbf{d}}_a(k,t)$ in the director frame (or body-fixed frame) are related to the unit vectors $\hat{\mathbf{a}}_a$ in the lab frame via an Euler rotation matrix $\Lambda$ according to
$$\hat{\mathbf{d}}_a(k,t) = \Lambda\big(\varphi(k,t),\theta(k,t),\psi(k,t)\big)\cdot\hat{\mathbf{a}}_a = \Lambda(k,t)\cdot\hat{\mathbf{a}}_a, \quad (a=1,2,3). \tag{2.1}$$
Since $\mathbf{r}\in R^3$ and $\Lambda\in SO(3)$, the configuration space of the elastic rod is $E^3 = R^3\times SO(3)$.

The orientation of the local frame at $s+\Delta s = (k+1)\Delta s$ is obtained by a tiny rotation of the coordinate frame at $s = k\Delta s$. The velocity of the rotation is the matrix
$$\overline{\Omega}(k,t) = (\Delta s)^{-1}\big[\Lambda(k,t)\big]^T\cdot\big[\Lambda(k+1,t)-\Lambda(k,t)\big]. \tag{2.2}$$
It can be shown that
$$\overline{\Omega}_{bc}(k,t) = \varepsilon_{abc}\Omega_a(k,t) + O(\Delta s) \tag{2.3}$$
where $\Omega_a(k,t)$ are the local components of the Darboux vector in the rod frame namely
$$\mathbf{\Omega}(k,t) = \Omega_a(k,t)\hat{\mathbf{d}}_a(k,t). \tag{2.4}$$
Here and after, double occurrence of an index in the subscript means summation over its range. Vectors $\hat{\mathbf{d}}_a(k,t)$ change with $k$ by moving perpendicular to themselves and to the rotation axis $\mathbf{\Omega}(k,t)$ according to
$$\hat{\mathbf{d}}_a(k+1,t) = \hat{\mathbf{d}}_a(k,t) + \Delta s\,\mathbf{\Omega}(k,t)\times\hat{\mathbf{d}}_a(k,t) + O(\Delta s^2). \tag{2.5}$$
The relative position of the origin of the localized rod frame at $s+\Delta s = (k+1)\Delta s$ is obtained by a tiny translation of the origin of the localized frame at $s = k\Delta s$, i.e.,
$$\mathbf{r}(k+1,t) = \mathbf{r}(k,t) + \Delta s\,\mathbf{\Gamma}(k,t) + O(\Delta s^2). \tag{2.6}$$
The velocity of the translation is the tangent vector $\mathbf{\Gamma}(k,t)$ with local components
$$\mathbf{\Gamma}(k,t) = \Gamma_a(k,t)\hat{\mathbf{d}}_a(k,t) \tag{2.7}$$
The parameter $s = k\Delta s$, usually chosen as the arclength parameter for the undeformed (or relaxed) elastic rod, is no longer the current arclength parameter for the deformed rod, $\tilde{s}(s,t)$, since there are deformations of shear and extension. The current arclength of the deformed rod, $\tilde{s}(s,t)$, is then given by $\tilde{s}(k,t) = \Delta s\sum_{n=1}^{k}|\mathbf{\Gamma}(n,t)|$.

The orientation of the local frame at time $t+\Delta t$ is obtained by an infinitesimal rotation of the coordinate frame at time $t$. The velocity of the rotation is the angular velocity vector $\boldsymbol{\omega}(k,t)$, with local frame components $\boldsymbol{\omega}(k,t) = \omega_a(k,t)\,\hat{\mathbf{d}}_a(k,t)$, obeying
$$\partial_t\hat{\mathbf{d}}_a(k,t) = \boldsymbol{\omega}(k,t)\times\hat{\mathbf{d}}_a(k,t). \tag{2.8}$$

The relative position of the origin of the localized rod frame at $t+\Delta t$ is obtained by an infinitesimal translation of the origin of the localized frame at $t$. The velocity of the translation is the vector

$$\partial_t \mathbf{r}(k,t) = \mathbf{\gamma}(k,t) = \gamma_a(k,t)\mathbf{d}_a(k,t). \tag{2.9}$$

Three dependent variables are used for describing strains (or deformations) of bending$_1$ ($\Omega_1$), bending$_2$ ($\Omega_2$), and twisting ($\Omega_3$). Another three are used for describing the strains (or deformations) of shear$_1$ ($\Gamma_1$), shear$_2$ ($\Gamma_2$), and extension ($\Gamma_3$). Still another three are used for describing the linear velocity$_1$ ($\gamma_1$), linear velocity$_2$ ($\gamma_2$), and linear velocity$_3$ ($\gamma_3$) for the translation of the centroid of the elastic rod cross section at position $s$. The last three are used for describing the angular velocity$_1$ ($\omega_1$), angular velocity$_2$ ($\omega_2$), and angular velocity$_3$ ($\omega_3$) for the rotation of the elastic rod cross section at position $s$.

At any time $t$ and a given position (say $s = s_1$) along the centerline, there is a cross section of the elastic rod upon which the internal forces are exerted. One side of the cross section ($s < s_1$) acts on the other side ($s > s_1$), and *vice versa*. The internal forces are resolved into a force $\mathbf{P}(s_1,t)$ and a torque $\mathbf{M}(s_1,t)$. At each cross section such a force and a torque may be found, giving rise to functions $\mathbf{P}(s,t)$ and $\mathbf{M}(s,t)$ describing a system of stresses on the elastic rod.

Since each cross section has its mass and moment of inertia tensor, an angular momentum $\mathbf{m}(s,t)$ of the cross section and linear velocity $\mathbf{p}(s,t)$ of the center of the cross section can be naturally introduced.

From this point on, the letters $\Omega, \Gamma, M, P, \omega, \gamma, m, p$ without subscripts will be understood as vectors, whether bold or not.

## III. LAX PAIR FOR THE s-DISCRETE SMK EQUATIONS

Let $J_a$ and $K_a$ ($a = 1,2,3$) be the generator matrices of Lie group $SO(4)$, given by

$$J_1 = \begin{pmatrix} 0 & 0 & 0 & 0 \\ 0 & 0 & -1 & 0 \\ 0 & 1 & 0 & 0 \\ 0 & 0 & 0 & 0 \end{pmatrix}, \quad J_2 = \begin{pmatrix} 0 & 0 & 1 & 0 \\ 0 & 0 & 0 & 0 \\ -1 & 0 & 0 & 0 \\ 0 & 0 & 0 & 0 \end{pmatrix}, \quad J_3 = \begin{pmatrix} 0 & -1 & 0 & 0 \\ 1 & 0 & 0 & 0 \\ 0 & 0 & 0 & 0 \\ 0 & 0 & 0 & 0 \end{pmatrix} \tag{3.1a}$$

$$K_1 = \begin{pmatrix} 0 & 0 & 0 & -1 \\ 0 & 0 & 0 & 0 \\ 0 & 0 & 0 & 0 \\ 1 & 0 & 0 & 0 \end{pmatrix}, \quad K_2 = \begin{pmatrix} 0 & 0 & 0 & 0 \\ 0 & 0 & 0 & -1 \\ 0 & 0 & 0 & 0 \\ 0 & 1 & 0 & 0 \end{pmatrix}, \quad K_3 = \begin{pmatrix} 0 & 0 & 0 & 0 \\ 0 & 0 & 0 & 0 \\ 0 & 0 & 0 & -1 \\ 0 & 0 & 1 & 0 \end{pmatrix}. \tag{3.1b}$$

These generators satisfy the relations

$$[J_a, J_b] = \varepsilon_{abc} J_c, \quad [J_a, K_b] = \varepsilon_{abc} K_c, \quad [K_a, K_b] = \varepsilon_{abc} J_c \tag{3.1c}$$

where $\varepsilon_{abc}$ is the Levi-Civita symbol. Now consider the linear system

$$\Phi(k+1, t, \lambda) = \tilde{U}(k, t, \lambda) \Phi(k, t, \lambda) \tag{3.2a}$$
$$\partial_t \Phi(k, t, \lambda) = \tilde{V}(k, t, \lambda) \Phi(k, t, \lambda) \tag{3.2b}$$

where $U$ and $V$ are defined as

$$\tilde{U} = \mathbf{1}_4 + \Delta s\, U \tag{3.3a}$$
$$\tilde{V} = V \tag{3.3b}$$
$$U = A + \sqrt{-1}\, B \tag{3.3c}$$
$$V = C + \sqrt{-1}\, D \tag{3.3d}$$

and where $\mathbf{1}_4$ is the 4-by-4 identity matrix and $A(k,t,\lambda)$, $B(k,t,\lambda)$, $C(k,t,\lambda)$, and $D(k,t,\lambda)$ are given by

$$A = -\left(\Omega_a J_a + \lambda^2 \Gamma_a K_a\right) \tag{3.4}$$
$$B = -\lambda \left(p_a J_a + \lambda^2 m_a K_a\right) \tag{3.5}$$
$$C = -\left(\omega_a J_a + \lambda^2 \gamma_a K_a\right) \tag{3.6}$$
$$D = -\lambda \left(P_a J_a + \lambda^2 M_a K_a\right). \tag{3.7}$$

In system (3.4-3.7), all symbols are real functions of $k$ and $t$ (except $\lambda$, the real spectral parameter).

In order to further simplify our presentation, we use the notation $f(k+1, t, \lambda) = f^1(k, t, \lambda) = f^1$.

The integrability condition for system (3.2) leads to the Lax equation for Lax pair $(\tilde{U}, \tilde{V})$:

$$\tfrac{d}{dt}\tilde{U} = \tilde{V}^1 \tilde{U} - \tilde{U} \tilde{V} \tag{3.8}$$

Substituting (3.3) into (3.8) and separating the real part from the imaginary part, we obtain:

$$\tfrac{d}{dt}\Gamma_c - (\Delta s)^{-1}\left(\gamma_c^1 - \gamma_c\right) \\ - \varepsilon_{abc}\left(\Gamma_b \omega_c^1 + \Omega_b \gamma_c\right) - \lambda^2 \varepsilon_{abc}\left(M_a p_b - m_a P_b^1\right) = 0 \tag{3.9a}$$

$$\frac{d}{dt}\Omega_a - (\Delta s)^{-1}\left(\omega_a^1 - \omega_a\right)$$
$$- \left(\Omega_{a+1}\,\omega_{a+2}^1 - \omega_{a+1}\,\Omega_{a+2}\right) - \lambda^2\left(P_{a+1}\,p_{a+2} - p_{a+1}\,P_{a+2}^1\right) \quad (3.9\text{b})$$
$$+ \lambda^4\left(\Gamma_{a+1}\,\gamma_{a+2} - \gamma_{a+1}^1\,\Gamma_{a+2}\right) + \lambda^6\left(M_{a+1}^1\,m_{a+2} - m_{a+1}\,M_{a+2}\right) = 0$$

$$\frac{d}{dt}p_a - (\Delta s)^{-1}\left(P_a^1 - P_a\right)$$
$$- \left(p_{a+1}\,\omega_{a+2}^1 - \omega_{a+1}\,p_{a+2}\right) + \left(P_{a+1}\,\Omega_{a+2} - \Omega_{a+1}\,P_{a+2}^1\right) \quad (3.9\text{c})$$
$$- \lambda^4\left(M_{a+1}\,\Gamma_{a+2} - \Gamma_{a+1}^1\,M_{a+2} + m_{a+1}\,\gamma_{a+2} - \gamma_{a+1}^1\,m_{a+2}\right) = 0$$

$$\frac{d}{dt}m_c - (\Delta s)^{-1}\left(M_c^1 - M_c\right)$$
$$- \varepsilon_{abc}\left(p_a\,\gamma_b + \Gamma_a\,P_b^1 + \Omega_a\,M_b + m_a\,\omega_b^1\right) = 0 \quad (3.9\text{d})$$

In (3.9b) and (3.9c) above, index $a$ runs 1,2,3. Index $a+1$ and index $a+2$ take their values of modulo 3.

In Appendix A we provide a different derivation of Eqs.(3.9a-d) using a different Lax pair.

When $\Delta s \to 0$, $k \to \infty$, system (3.9) reduces to the s-and-t continuous counterpart [(3.5) in Paper I].

Since system (3.9a-d) contains 12 scalar equations for 24 real dependent variables, $\Omega_a, \Gamma_a, \omega_a, \gamma_a, M_a, P_a, \omega_a, p_a$ ($a = 1,2,3$) we have the freedom to pick 12 real "constitutive" relations. In elastic work one uses the 12 scalar equations implied by the four vector equations

$$P_a = \frac{\partial}{\partial \Gamma_a} H(\Omega, \Gamma) \quad (3.10\text{a})$$

$$M_a = \frac{\partial}{\partial \Omega_a} H(\Omega, \Gamma) \quad (3.10\text{b})$$

$$p_a = \frac{\partial}{\partial \gamma_a} h(\omega, \gamma, t) \quad (3.10\text{c})$$

$$m_a = \frac{\partial}{\partial \omega_a} h(\omega, \gamma, t) \quad (3.10\text{d})$$

where ($a = 1,2,3$) and $H(\Omega, \Gamma)$ is the elastic energy function and $h(\omega, \gamma, t)$ is the kinetic energy function. But in other problems one might use other relations, which we write symbolically as

$$H_A(\Omega, \Gamma, M, P, \omega, \gamma, m, p, t, \lambda) = 0, \qquad (A = 1, 2, \ldots, 12). \quad (3.11)$$

System (3.9) and (3.11) can describe a very large class of differential-difference equations [24 dependent variables in (1+1) dimension]. Some members of this class will be discussed in detail in Section IV, below.

In Paper I we discussed the similarities and differences between the Lax pair formulation of the elastic rod problem and the more general Lax pair formulation of Doliwa and Santini[3] for the kinematics of a moving space curve of any kind. The same ideas apply when the problem is s-discretized, but the discussion differs enough in detail that it bears repeating.

The pure kinematic approach for recasting the nonlinear differential-difference equations of a general curve into Lax representation focuses only on what we call the strain-velocity compatibility (integrability) relations for equations like

$$\Phi^1 = Q(\Omega, \Gamma, t, \lambda)\,\Phi, \qquad \text{and} \qquad \partial_t \Phi = R(\omega, \gamma, t, \lambda)\,\Phi$$

where $\Omega$ and $\Gamma$ are strains and $\omega$ and $\gamma$ are velocities. But we treat strain-velocity and stress-momentum on an equal footing, so our equations are

$$\Phi^1 = U(\Omega, \Gamma, p, m, t, \lambda)\,\Phi, \quad \text{and} \qquad \partial_t \Phi = V(\omega, \gamma, P, M, t, \lambda)\,\Phi$$

where the new variables $P$ and $M$ are stresses (force and torque, respectively) and $p$ and $m$ are momenta (linear and angular, respectively). The Lax equations for the two cases look identical

$$\tfrac{d}{dt} Q = R^1 Q - Q R \qquad \text{versus} \qquad \tfrac{d}{dt} U = V^1 U - U V,$$

and quantities $Q$, $R$, $U$, and $V$ are all 4-by-4 arrays, but $Q$ and $R$ are real, whereas our $U$ and $V$ are complex and contain twice as many dependent variables.

Expanding the dependent variables $X_a = X_a(t,\lambda)$ and the constitutive relations $H_A(\Omega, \Gamma, M, P, \omega, \gamma, m, p, t, \lambda)$ in Taylor series in $\lambda$,

$$X_a(t,\lambda) = \sum_{n=0}^{\infty} \lambda^n X_a^{(n)}(t) \text{ and } H_A(t,\lambda) = \sum_{n=0}^{\infty} \lambda^n H_A^{(n)}(t), \qquad (3.12)$$

and taking the limit $\lambda \to 0$, we find that the leading terms in the expansions (3.9) are the s-discrete (t-continuous) SMK equations:

$$\tfrac{d}{dt}\Gamma_c - (\Delta s)^{-1}\left(\gamma_c^1 - \gamma_c\right) \\ \quad - \varepsilon_{abc}\left(\Gamma_b\,\omega_c^1 + \Omega_b\,\gamma_c\right) = 0 \qquad (3.13a)$$

$$\tfrac{d}{dt}\Omega_a - (\Delta s)^{-1}\left(\omega_a^1 - \omega_a\right) - \left(\Omega_{a+1}\,\omega_{a+2}^1 - \omega_{a+1}\,\Omega_{a+2}\right) = 0 \qquad (3.13b)$$

$$\frac{d}{dt} p_a - (\Delta s)^{-1} \left( P_a^1 - P_a \right)$$
$$- \left( p_{a+1} \omega_{a+2}^1 - \omega_{a+1} p_{a+2} \right) + \left( P_{a+1} \Omega_{a+2} - \Omega_{a+1} P_{a+2}^1 \right) = 0 \quad (3.13c)$$

$$\frac{d}{dt} m_c - (\Delta s)^{-1} \left( M_c^1 - M_c \right)$$
$$- \varepsilon_{abc} \left( p_a \gamma_b + \Gamma_a P_b^1 + \Omega_a M_b + m_a \omega_b^1 \right) = 0 \quad (3.13d)$$

In (3.13b) and (3.13c) above, index $a$ runs 1,2,3. Index $a+1$ and index $a+2$ take their values of modulo 3. When $\Delta s \to 0$ the s-discrete SMK equations reduce to s-continuous SMK equations ((2.1) in Paper I).

## IV. SPECIALIZATION OF THE SMK EQUATIONS

The great generality of the elastic rod problem provides many specialized cases which have generated entire literatures on their own. To emphasize that these now all belong to the same class of integrable problems, we now discuss several famous sub-problems in detail, showing exactly how they relate to the general formulation presented above.

### Case 1. The Static Elastic Rod

Setting velocities and momenta $\omega_a = \gamma_a = m_a = p_a = 0$ in (3.13) and assuming that everything else is independent variable $t$, we find that the s-discrete SMK equations (3.13) reduce to the following difference equations:

$$(\Delta s)^{-1} \left( M_c^1 - M_c \right) + \varepsilon_{abc} \left( \Gamma_a P_b^1 + \Omega_a M_b \right) = 0 \quad (4.1a)$$

$$(\Delta s)^{-1} \left( P_a^1 - P_a \right) - \left( P_{a+1} \Omega_{a+2} - \Omega_{a+1} P_{a+2}^1 \right) = 0 \quad (4.1b)$$

$$P_a = \frac{\partial}{\partial \Gamma_a} H(\Omega, \Gamma) \quad (4.1c)$$

$$M_a = \frac{\partial}{\partial \Omega_a} H(\Omega, \Gamma). \quad (4.1d)$$

System (4.1) describes the equilibrium configurations of an elastic rod with elastic energy function $H(\Omega, \Gamma)$.

### Subcase 1.1  Static Elastic Rod with Linear Constitutive Relations

Let $H(\Omega, \Gamma)$ of (4.1c) and (4.1d) be given by

$$H(\Omega, \Gamma) = \tfrac{1}{2} A_{ab} \left( \Omega_a - \Omega_a^{(\text{intrinsic})} \right) \left( \Omega_b - \Omega_b^{(\text{intrinsic})} \right)$$
$$+ \tfrac{1}{2} C_{ab} \left( \Gamma_a - \Gamma_a^{(\text{intrinsic})} \right) \left( \Gamma_b - \Gamma_b^{(\text{intrinsic})} \right)$$
$$+ \tfrac{1}{2} B_{ab} \left[ \left( \Omega_a - \Omega_a^{(\text{intrinsic})} \right) \left( \Gamma_b - \Gamma_b^{(\text{intrinsic})} \right) \right.$$
$$\left. + \left( \Gamma_a - \Gamma_a^{(\text{intrinsic})} \right) \left( \Omega_b - \Omega_b^{(\text{intrinsic})} \right) \right] \quad (4.2)$$

where $A_{ab}$ is the bending/twisting modulus, $C_{ab}$ is the shear/extension modulus, and $B_{ab}$ is the coupling modulus between bending/twisting and shear/extension. The quantities $\Omega_a^{(\text{intrinsic})}$ are the intrinsic bending and twisting of the unstressed rod, and the quantities $\Gamma_a^{(\text{intrinsic})}$ are the intrinsic shear and extension. Then system (4.1) reduces to

$$(\Delta s)^{-1} \left( M_c^1 - M_c \right) + \varepsilon_{abc} \left( \Gamma_a P_b^1 + \Omega_a M_b \right) = 0 \quad (4.3a)$$
$$(\Delta s)^{-1} \left( P_a^1 - P_a \right) - \left( P_{a+1} \Omega_{a+2} - \Omega_{a+1} P_{a+2}^1 \right) = 0 \quad (4.3b)$$
$$M_a = A_{ab} \left( \Omega_b - \Omega_b^{(\text{intrinsic})} \right) + B_{ab} \left( \Gamma_b - \Gamma_b^{(\text{intrinsic})} \right) \quad (4.3c)$$
$$P_a = C_{ab} \left( \Gamma_b - \Gamma_b^{(\text{intrinsic})} \right) + B_{ab} \left( \Omega_b - \Omega_b^{(\text{intrinsic})} \right). \quad (4.3d)$$

The s-continuous limit of the system (4.3a,b) in this paper is identical to system (4.3a,b) in paper I. We have shown in paper I that, when
(1) $B_{ab} = 0$ and
(2) $A_{ab}$ is a constant diagonal matrix with $A_{11} = A_{22}$, and
(3) $C_{ab}$ is a constant diagonal matrix with $C_{11} = C_{22}$, and
(4) $\Omega_a^{(\text{intrinsic})} = 0$, $\Gamma_1^{(\text{intrinsic})} = \Gamma_2^{(\text{intrinsic})} = \Gamma_3^{(\text{intrinsic})} - 1 = 0$,
the s-continuous limit of the system (4.3a,b) may be solved exactly in terms of elliptic functions[4]. Open question: When the above conditions pertain, is this difference system (4.3a,b) also integrable?

### Subcase 1.2   Static Kirchhoff Elastic Rod, and the Heavy Top

Assuming that elastic rod does not have shear and extension deformations [*i.e.*, $\Gamma = \Gamma^{(\text{intrinsic})} = (0, 0, 1)^T$], then we find that $H(\Omega)$ becomes

$$H(\Omega) = \tfrac{1}{2} A_{ab} \left( \Omega_a - \Omega_a^{(\text{intrinsic})} \right) \left( \Omega_b - \Omega_b^{(\text{intrinsic})} \right), \quad (4.4a)$$

and system (4.3) reduces to

$$(\Delta s)^{-1} \left( M_c^1 - M_c \right) + \varepsilon_{abc} \left( \Gamma_a P_b^1 + \Omega_a M_b \right) = 0 \quad (4.5a)$$
$$(\Delta s)^{-1} \left( P_a^1 - P_a \right) - \left( P_{a+1} \Omega_{a+2} - \Omega_{a+1} P_{a+2}^1 \right) = 0 \quad (4.5b)$$

$$M_a = A_{ab}\left(\Omega_b - \Omega_b^{(\text{intrinsic})}\right). \tag{4.5c}$$

If $s = k\Delta s$ is arc length (as assumed earlier), then system (4.5) describes the equilibrium configuration of the unshearable, inextensible discrete Kirchhoff elastic rod. But if $s = k\Delta s$ is understood as time, this system describes the time-discrete dynamics of the heavy top.

There are two known integrable cases for the time-continuous version of the heavy top system (4.5) with $\Omega^{(\text{intrinsic})} = (0,0,0)^T$, $\Gamma = (0,0,1)^T$:

(a) Lagrange Top[5]: $A_{ab}$ is a constant diagonal matrix with $A_{11} = A_{22}$.
(b) Kowalewski Top[6]: the same, but with $A_{33} = A_{11} = 2A_{22}$.

Open question: what about the time-discrete dynamics of the Lagrange top and the Kowalewski top?

## Case 2, Rigid Body Motion In Ideal Fluid

Setting $M = P = \Omega = \Gamma = (0,0,0)^T$ in (3.13) and assuming that everything else is a function of time $t$ only, then we find that the SMK equations (3.13) reduce to

$$\tfrac{d}{dt} p + \omega \times p = 0 \tag{5.1a}$$

$$\tfrac{d}{ds} m + \omega \times m + \gamma \times p = 0 \tag{5.1b}$$

$$\gamma_a = \frac{\partial}{\partial p_a} h(m, p, t) \tag{5.1c}$$

$$\omega_a = \frac{\partial}{\partial m_a} h(m, p, t). \tag{5.1d}$$

This case has been discussed in paper I.

## Case 3, Kirchhoff Elastic Rod Motion

In this case, there is no shear or extension, so $\Gamma = \Gamma^{(\text{intrinsic})} = (0, 0, 1)^T$ is independent of time t. SMK equations in (3.13) reduce to:

$$\begin{aligned}&\tfrac{d}{dt} m_c - (\Delta s)^{-1}\left(M_c^1 - M_c\right) \\ &- \varepsilon_{abc}\left(p_a \gamma_b + \Gamma_a P_b^1 + \Omega_a M_b + m_a \omega_b^1\right) = 0\end{aligned} \tag{6.2a}$$

$$\tfrac{d}{dt}\Omega_a - (\Delta s)^{-1}\left(\omega_a^1 - \omega_a\right) - \left(\Omega_{a+1} \omega_{a+2}^1 - \omega_{a+1} \Omega_{a+2}\right) = 0 \tag{6.2b}$$

$$\begin{aligned}&\tfrac{d}{dt} p_a - (\Delta s)^{-1}\left(P_a^1 - P_a\right) \\ &- \left(p_{a+1} \omega_{a+2}^1 - \omega_{a+1} p_{a+2}\right) + \left(P_{a+1} \Omega_{a+2} - \Omega_{a+1} P_{a+2}^1\right) = 0\end{aligned} \tag{6.2c}$$

System (6.2) describes the dynamics of the Kirchhoff elastic rod if we pick

$$M_a = A_{ab}\left(\Omega_b - \Omega_b^{(\text{intrinsic})}\right) \quad (6.2e)$$

$$m_a = I_{ab}\,\omega_b \quad (6.2f)$$

$$p_a = \rho\,\gamma_a \quad (6.2g)$$

where $I_{ab}$ is the moment of inertia tensor for the cross section of the elastic rod and $\rho$ is the linear density of mass.

## Case 4, Elastic Rod Moving in a Plane

Setting $\omega_1 = \omega_2 = m_1 = m_2 = \gamma_3 = p_3 = 0$ and $\Omega_1 = \Omega_2 = M_1 = M_2 = \Gamma_3 = P_3 = 0$ in (3.13), then we find that the s-discrete SMK equations reduce to:

$$\tfrac{d}{dt}\Omega_3 - (\Delta s)^{-1}\left(\omega_3^1 - \omega_3\right) = 0 \quad (7.1a)$$

$$\tfrac{d}{dt}\Gamma_1 - (\Delta s)^{-1}\left(\gamma_1^1 - \gamma_1\right) - \Gamma_2\,\omega_3^1 + \gamma_2\,\Omega_3 = 0 \quad (7.1b)$$

$$\tfrac{d}{dt}\Gamma_2 - (\Delta s)^{-1}\left(\gamma_2^1 - \gamma_2\right) + \Gamma_1\,\omega_3^1 - \gamma_1\,\Omega_3 = 0 \quad (7.1c)$$

$$\tfrac{d}{dt}m_3 - (\Delta s)^{-1}\left(M_3^1 - M_3\right) - p_1\,\gamma_2 + p_2\,\gamma_1 - \Gamma_1\,P_2^1 + \Gamma_2\,P_1^1 = 0 \quad (7.1d)$$

$$\tfrac{d}{dt}p_1 - (\Delta s)^{-1}\left(P_1^1 - P_1\right) - p_2\,\omega_3^1 + P_2\,\Omega_3 = 0 \quad (7.1e)$$

$$\tfrac{d}{dt}p_2 - (\Delta s)^{-1}\left(P_2^1 - P_2\right) + p_1\,\omega_3^1 - P_1\,\Omega_3 = 0. \quad (7.1f)$$

System (7.1a – 7.1f) contains six scalar equations for twelve dependent variables. The six constitutive relations can be expressed as

$$P_1 = \frac{\partial}{\partial \Gamma_1} H(\Omega_3, \Gamma_1, \Gamma_2) \quad (7.2a)$$

$$P_2 = \frac{\partial}{\partial \Gamma_2^k} H(\Omega_3, \Gamma_1, \Gamma_2) \quad (7.2b)$$

$$M_3 = \frac{\partial}{\partial \Omega_3} H(\Omega_3, \Gamma_1, \Gamma_2) \quad (7.2c)$$

$$p_1 = \frac{\partial}{\partial \gamma_1} h(\omega_3, \gamma_1, \gamma_2) \quad (7.2d)$$

$$p_2 = \frac{\partial}{\partial \gamma_2} h(\omega_3, \gamma_1, \gamma_2) \tag{7.2e}$$

$$m_3 = \frac{\partial}{\partial \omega_3} h(\omega_3, \gamma_1, \gamma_2). \tag{7.2f}$$

In system (7.1), the dependent variable $\Omega_3$ is used for describing bending, $\Gamma_1$ for shear, $\Gamma_2$ for extension, $\gamma_1$ for linear velocity₁, $\gamma_2$ for linear velocity₂, and $\omega_3$ for angular velocity. In this system, $H(\Omega_3, \Gamma_1, \Gamma_2)$ stands for the elastic energy and $h(\omega_3, \gamma_1, \gamma_2)$ for the kinetic energy.

## Case 5, Moving Space Curve (3D)

The first two kinematic vector equations (3.13a, 3.13b) in the SMK equations (3.13) are derived from the following Serret-Frenet equations for an elastic rod moving on $S^3$ with radius $\lambda^{-2}$, namely,

$$\begin{pmatrix} \hat{\mathbf{d}}_1(k+1,t) \\ \hat{\mathbf{d}}_2(k+1,t) \\ \hat{\mathbf{d}}_3(k+1,t) \\ \hat{\mathbf{r}}(k+1,t) \end{pmatrix} = (\Delta s) \begin{pmatrix} (\Delta s)^{-1} & \Omega_3 & -\Omega_2 & -\lambda^2 \Gamma_1 \\ -\Omega_3 & (\Delta s)^{-1} & -\Omega_1 & -\lambda^2 \Gamma_2 \\ -\Omega_2 & \Omega_1 & (\Delta s)^{-1} & -\lambda^2 \Gamma_3 \\ \lambda^2 \Gamma_1 & \lambda^2 \Gamma_2 & \lambda^2 \Gamma_3 & (\Delta s)^{-1} \end{pmatrix} \begin{pmatrix} \hat{\mathbf{d}}_1(k,t) \\ \hat{\mathbf{d}}_2(k,t) \\ \hat{\mathbf{d}}_3(k,t) \\ \hat{\mathbf{r}}(k,t) \end{pmatrix} \tag{8.1a}$$

$$\frac{d}{dt} \begin{pmatrix} \hat{\mathbf{d}}_1(k,t) \\ \hat{\mathbf{d}}_2(k,t) \\ \hat{\mathbf{d}}_3(k,t) \\ \hat{\mathbf{r}}(k,t) \end{pmatrix} = \begin{pmatrix} 0 & \omega_3 & -\omega_2 & -\lambda^2 \gamma_1 \\ -\omega_3 & 0 & -\omega_1 & -\lambda^2 \gamma_2 \\ -\omega_2 & \omega_1 & 0 & -\lambda^2 \gamma_3 \\ \lambda^2 \gamma_1 & \lambda^2 \gamma_2 & \lambda^2 \gamma_3 & 0 \end{pmatrix} \begin{pmatrix} \hat{\mathbf{d}}_1(k,t) \\ \hat{\mathbf{d}}_2(k,t) \\ \hat{\mathbf{d}}_3(k,t) \\ \hat{\mathbf{r}}(k,t) \end{pmatrix}. \tag{8.1b}$$

Eq. (8.1a) can be obtained from Eqs. (2.5) and (2.6) when the $O(\Delta s^2)$ is omitted. Eq. (8.1b) can be obtained similarly from Eqs. (2.8) and (2.9).

If we assume zero elastic energy $H(\Omega, \Gamma)$ and zero kinetic energy $h(\omega, \gamma, t)$ in the constitutive relations (3.10a-d), then $P = M = p = m = (0,0,0)^T$. Consequently the last two dynamic vector equations (3.13c, 3.13d) in the SMK equations (3.13) for force balance and torque balance become zero identically.

Furthermore, if we let
$$\hat{\mathbf{d}}_3 = \hat{\mathbf{t}}, \ \hat{\mathbf{d}}_2 = \hat{\mathbf{n}}, \ \hat{\mathbf{d}}_1 = \hat{\mathbf{b}} \tag{8.2a}$$
$$\kappa = \Omega_1/\Gamma_3, \ \tau = -\Omega_3/\Gamma_3. \tag{8.2b}$$
$$\Omega_2 = \Gamma_1 = \Gamma_2 = 0, \tag{8.2c}$$
then (8.1a) and (8.1b) become

$$\begin{pmatrix} \hat{\mathbf{b}}(k+1,t) \\ \hat{\mathbf{n}}(k+1,t) \\ \hat{\mathbf{t}}(k+1,t) \\ \hat{\mathbf{r}}(k+1,t) \end{pmatrix} = (\Delta\tilde{s}) \begin{pmatrix} (\Delta\tilde{s})^{-1} & -\tau & 0 & 0 \\ \tau & (\Delta\tilde{s})^{-1} & -\kappa & 0 \\ 0 & \kappa & (\Delta\tilde{s})^{-1} & -\lambda^2 \\ 0 & 0 & \lambda^2 & (\Delta\tilde{s})^{-1} \end{pmatrix} \begin{pmatrix} \hat{\mathbf{b}}(k,t) \\ \hat{\mathbf{n}}(k,t) \\ \hat{\mathbf{t}}(k,t) \\ \hat{\mathbf{r}}(k,t) \end{pmatrix} \quad (8.3a)$$

$$\frac{d}{dt}\begin{pmatrix} \hat{\mathbf{d}}_1(k,t) \\ \hat{\mathbf{d}}_2(k,t) \\ \hat{\mathbf{d}}_3(k,t) \\ \hat{\mathbf{r}}(k,t) \end{pmatrix} = \begin{pmatrix} 0 & \omega_3 & -\omega_2 & -\lambda^2\gamma_1 \\ -\omega_3 & 0 & -\omega_1 & -\lambda^2\gamma_2 \\ -\omega_2 & \omega_1 & 0 & -\lambda^2\gamma_3 \\ \lambda^2\gamma_1 & \lambda^2\gamma_2 & \lambda^2\gamma_3 & 0 \end{pmatrix} \begin{pmatrix} \hat{\mathbf{d}}_1(k,t) \\ \hat{\mathbf{d}}_2(k,t) \\ \hat{\mathbf{d}}_3(k,t) \\ \hat{\mathbf{r}}(k,t) \end{pmatrix} \quad (8.3b)$$

where $\Delta\tilde{s} = \Gamma_3 \Delta s$.

With $\tilde{s} = k\Delta\tilde{s}$ Eqs. (8.3a) and (8.3b) become the relations satisfied by the Serret-Frenet frame $\{\hat{\mathbf{n}}(\tilde{s},t,\lambda), \hat{\mathbf{b}}(\tilde{s},t,\lambda), \hat{\mathbf{t}}(\tilde{s},t,\lambda), \hat{\mathbf{r}}(\tilde{s},t,\lambda)\}$ when a discrete space curve moves on a real 3-dimensional spherical surface $S^3$ with radius $\lambda^{-2}$. In Eqs. (8.3a) and (8.3b), $\hat{\mathbf{r}}(\tilde{s},t,\lambda) = \lambda^{-2}\mathbf{r}(\tilde{s},t,\lambda)$ is the unit radius vector, $\kappa = \kappa(\tilde{s},t,\lambda)$ is the curvature, and $\tau = \tau(\tilde{s},t,\lambda)$ is the geometric torsion of the discrete space curve.

Since $\Gamma_3$ is a function of $(k,t)$, then $\Delta\tilde{s}$, the step size in arclength $\tilde{s}$, is a function of $(k,t)$ as well.

When $\Gamma_3$ becomes a constant, $\Delta\tilde{s}$ also becomes a constant as well. Doliwa and Santini have investigated this situation[3].

Thus a discrete space curve with constant step size in arclength is a special case of the discrete elastic rod with no shear deformation ($\Gamma_1 = \Gamma_2 = 0$), with constant extension ($\Gamma_3 = \text{constant}$), with no bending deformation in $\hat{\mathbf{d}}_2 = \hat{\mathbf{n}}$ direction ($\Omega_2 = 0$), with zero elastic energy function $H(\Omega,\Gamma)$, and with zero kinetic energy function $h(\omega,\gamma)$.

This is the key link connecting the s-discrete (and t-continuous) SMK equations (3.13), via the motion of a discrete space curve problem (8.3a,b), with most well-known integrable discrete systems in dimension (1+1) (Ablowitz-Ladik etc) [7].

There also exists another tetrad frame $(\hat{\mathbf{e}}_1(s,t,\lambda), \hat{\mathbf{e}}_2(s,t,\lambda), \hat{\mathbf{e}}_3(s,t,\lambda), \hat{\mathbf{e}}_4(s,t,\lambda))$ which satisfies the following relations:

$$\begin{pmatrix} \hat{\mathbf{e}}_1(k+1,t) \\ \hat{\mathbf{e}}_2(k+1,t) \\ \hat{\mathbf{e}}_3(k+1,t) \\ \hat{\mathbf{e}}_4(k+1,t) \end{pmatrix} = (\Delta s) \begin{pmatrix} (\Delta s)^{-1} & p_3 & -p_2 & -\lambda^2 \, m_1 \\ -p_3 & (\Delta s)^{-1} & -p_1 & -\lambda^2 \, m_2 \\ -p_2 & p_1 & (\Delta s)^{-1} & -\lambda^2 \, m_3 \\ \lambda^2 \, m_1 & \lambda^2 \, m_2 & \lambda^2 \, m_3 & (\Delta s)^{-1} \end{pmatrix} \begin{pmatrix} \hat{\mathbf{e}}_1(k,t) \\ \hat{\mathbf{e}}_2(k,t) \\ \hat{\mathbf{e}}_3(k,t) \\ \hat{\mathbf{e}}_4(k,t) \end{pmatrix} \quad (8.4a)$$

$$\frac{d}{dt}\begin{pmatrix} \hat{\mathbf{e}}_1(k,t) \\ \hat{\mathbf{e}}_2(k,t) \\ \hat{\mathbf{e}}_3(k,t) \\ \hat{\mathbf{e}}_4(k,t) \end{pmatrix} = \begin{pmatrix} 0 & P_3 & -P_2 & -\lambda^2 \, M_1 \\ -P_3 & 0 & -P_1 & -\lambda^2 \, M_2 \\ -P_2 & P_1 & 0 & -\lambda^2 \, M_3 \\ \lambda^2 \, M_1 & \lambda^2 \, M_2 & \lambda^2 \, M_3 & 0 \end{pmatrix} \begin{pmatrix} \hat{\mathbf{e}}_1(k,t) \\ \hat{\mathbf{e}}_2(k,t) \\ \hat{\mathbf{e}}_3(k,t) \\ \hat{\mathbf{e}}_4(k,t) \end{pmatrix}. \quad (8.4b)$$

Since the elements in 4-by-4 matrices in (8.2a,b) are related to the elements of the 4-by-4 matrices in (8.4a,b) via the constitutive relations like (3.10a-d), we may say that the frame $\{\hat{\mathbf{e}}_1, \hat{\mathbf{e}}_2, \hat{\mathbf{e}}_3, \hat{\mathbf{e}}_4\}$ is dual to the Serret-Frenet frame $\{\tilde{\mathbf{d}}_1, \tilde{\mathbf{d}}_2, \tilde{\mathbf{d}}_3, \hat{\mathbf{r}}\}$.

## V. RELATION TO SPIN DESCRIPTION OF SOLITON EQUATIONS

We may rewrite the first component equation in (8.2b) as

$$\hat{\mathbf{S}}(k+1,t) = \hat{\mathbf{S}}(k,t) + \Delta s \sum_{J=2}^{4} b_J \, \hat{\mathbf{d}}_J(k,t), \qquad \hat{\mathbf{S}}(k,t) \equiv \hat{\mathbf{d}}_1(k,t) \quad (9.1)$$

This is a discrete version of the basic equation in Myrzakulov's unit spin description of the integrable and nonintegrable PDEs [8].

Eqs. (8.1a,b) tell us that it is better to consider the motion of not just unit vector $\hat{\mathbf{S}} = \hat{\mathbf{d}}_1$, but rather the motion of all unit vectors $\hat{\mathbf{d}}_J$ ($J = 1,2,3,4$) together in (1+1) dimension. We conjecture that any system of integrable or nonintegrable Differential Difference Equations in (1+1) dimension derived from Eq. (9.1) might also be derived from a system of Differential Difference Equations in (3.9) with Lax pair $(\tilde{U}, \tilde{V})$ of (3.2) or Lax pair $(\tilde{X}, \tilde{Y})$ of (A.1).

If we use the 8-by-8 Lax pair $(\tilde{X}, \tilde{Y})$ as shown in Appendix A and choose the normalization factor properly for $\Psi$ in (A.1), then we can rewrite (A.1) as

$$\hat{\mathbf{f}}_I(k+1,t) = \hat{\mathbf{f}}_I(k,t) + \Delta s \, X_{IJ}(k,t) \, \hat{\mathbf{f}}_J(k,t) \qquad (I = 1,2,\ldots,8) \quad (9.2a)$$

$$\tfrac{d}{dt}\hat{\mathbf{f}}_I(k,t) = Y_{IJ}(k,t) \, \hat{\mathbf{f}}_J(k,t), \qquad (I = 1,2,\ldots,8) \quad (9.2b)$$

where $(\hat{\mathbf{f}}_1, \hat{\mathbf{f}}_2, \ldots, \hat{\mathbf{f}}_8)^T = \Psi^T$, $\hat{\mathbf{f}}_I \cdot \hat{\mathbf{f}}_J = \delta_{IJ}$, and $X$ and $Y$ are the 8-by-8 matrices defined in (A.2a) and (A.2b) respectively.

Matrices $X$ and $Y$ have the following symmetry properties:

$$X_{IJ} = X_{JI}, \qquad Y_{IJ} = Y_{JI} \qquad \text{(if } I,J-4 = 1,2,3,4 \text{ or } I-4, J = 1,2,3,4\text{)} \quad (9.3a)$$

$$X_{IJ} = -X_{JI}, \qquad Y_{IJ} = -Y_{JI} \qquad \text{(if } I,J = 1,2,3,4 \text{ or } I,J = 5,6,7,8\text{)}. \quad (9.3b)$$

Since matrix $X$ is not antisymmetric, Eq. (9.2a) is not a Serret-Frenet equation for an elastic rod moving on a 7D sphere ($S^7$) with radius $|\xi|^{-2} = \lambda^{-2}$ imbedded in $R^8$.

Because the diagonal matrix elements of $Y$ are all zero, we may rewrite the first component equation in (9.2b) as

$$\hat{\mathbf{S}}(k+1,t) = \hat{\mathbf{S}}(k,t) + \Delta s \sum_{J=2}^{8} a_J(k,t)\,\hat{\mathbf{f}}_J(k,t), \qquad \hat{\mathbf{S}}(k,t) \equiv \hat{\mathbf{f}}_1(k,t) \tag{9.4}$$

where $a_J(k,t) \equiv Y_{1J}(k,t)$.

This is the discrete analog of Eq. (9.1) in 8D[8]. Again Eq. (9.2) tell us that it is better to consider the motion of not just unit vector $\hat{\mathbf{S}} \equiv \hat{\mathbf{f}}_1$, but rather the motion of all unit vectors $\hat{\mathbf{f}}_J$ ($J = 1,2,...,7,8$) together in (1+1) dimension.

## VI. CONCLUSIONS

(1) We have found a Lax pair with a corresponding spectral parameter for a system of 12 scalar Differential-Difference-Equations and 12 scalar constitutive relations governing 24 dependent variables in (1+1) dimension.

(2) When the spectral parameter goes to zero, this system of Differential-Difference-Equations reduces to the s-discrete SMK equations that describe the dynamics of DNA modeled as a shearable and extensible discrete elastic rod.

(3) When three dependent variables are set to be constants, the s-discrete SMK equations reduce to a system of 9 scalar Differential-Difference-Equations with 9 scalar constitutive relations in 18 dependent variables. This system describes the dynamics of DNA modeled as an unshearable and inextensible discrete elastic rod (discete Kirchhoff elastic rod).

(4) When the s-discrete SMK equations are assumed to be independent of t, they reduce to a set of 6 Difference-Equations and 3 constitutive relations for 9 dependent variables describing the discrete motion of a heavy top.

(5) When the s-discrete SMK equations are assumed to be independent of k, they reduce to a set of 6 ODEs and 3 constitutive relations for 9 dependent variables describing the motion of a rigid body in an ideal fluid.

## APPENDIX A

If we assume that 24 dependent variables $\Omega_a, \Gamma_a, \omega_a, \gamma_a, M_a, P_a, m_a, p_a$ ($a=1,2,3$) are complex and the parameter $\zeta$ is also complex, then we may consider the following linear system:

$$\Psi^1 = \tilde{X}\,\Psi \tag{A.1a}$$

$$\tfrac{d}{dt}\Psi = \tilde{Y}\,\Psi \tag{A.1b}$$

where $\tilde{X}$ and $\tilde{Y}$ in the Lax pair $(\tilde{X},\tilde{Y})$ are defined as

$$\tilde{X} = \mathbf{1}_8 + \Delta s\, X \tag{A.2a}$$

$$\tilde{Y} = Y \tag{A.2b}$$

$$X = \begin{pmatrix} A & -B \\ B & A \end{pmatrix} \tag{A.2c}$$

$$Y = \begin{pmatrix} C & -D \\ D & C \end{pmatrix} \tag{A.2d}$$

and where $\mathbf{1}_8$ is the 8-by-8 unit matrix and $A, B, C, D$ are given by

$$A = -\Omega_a\, J_a + \zeta^2\, \Gamma_a\, K_a \tag{A.2e}$$

$$B = \zeta\left(-p_a\, J_a + \zeta^2\, m_a\, K_a\right) \tag{A.2f}$$

$$C = -\omega_a\, J_a + \zeta^2\, \gamma_a\, K_a \tag{A.2g}$$

$$D = \zeta\left(-P_a\, J_a + \zeta^2\, M_a\, K_a\right) \tag{A.2h}$$

The integrability condition for system (A.1) leads to:

$$\tfrac{d}{dt}\tilde{X} = \tilde{Y}^1\,\tilde{X} - \tilde{X}\,\tilde{Y} \tag{A.3}$$

Taking each 4-by-4 block in (A.3) and left- or right-multiplying it by $J_a$ and $K_a$ ($a=1,2,3$) respectively, using (3.1c) in the main text to simplify the result, we obtain a set of Differential Difference Equations:

$$\begin{aligned}
&\tfrac{d}{dt}\Gamma_c - (\Delta s)^{-1}\left(\gamma_c^1 - \gamma_c\right) \\
&\quad -\varepsilon_{abc}\left(\Gamma_a\, \omega_b^1 + \Omega_a\, \gamma_b\right) + (\zeta)^2\,\varepsilon_{abc}\left(M_a\, p_b - m_a\, P_b^1\right) = 0
\end{aligned} \tag{A.4a}$$

$$\begin{aligned}
&\tfrac{d}{dt}\Omega_a - (\Delta s)^{-1}\left(\omega_a^1 - \omega_a\right) \\
&\quad -\left(\Omega_{a+1}\, \omega_{a+2}^1 - \omega_{a+1}\,\Omega_{a+2}\right) + (\zeta)^2\left(P_{a+1}\, p_{a+2} - p_{a+1}\, P_{a+2}^1\right) \\
&\quad +(\zeta)^4\left(\Gamma_{a+1}\,\gamma_{a+2} - \gamma_{a+1}^1\,\Gamma_{a+2}\right) - (\zeta)^6\left(M_{a+1}^1\, m_{a+2} - m_{a+1}\, M_{a+2}\right) = 0
\end{aligned} \tag{A.4b}$$

$$\begin{aligned}
&\tfrac{d}{dt}p_a - (\Delta s)^{-1}\left(P_a^1 - P_a\right) \\
&\quad -\left(p_{a+1}\,\omega_{a+2}^1 - \omega_{a+1}\, p_{a+2}\right) + \left(P_{a+1}\,\Omega_{a+2} - \Omega_{a+1}\, P_{a+2}^1\right) \\
&\quad +(\zeta)^4\left(\Gamma_{a+1}\, M_{a+2}^1 - M_{a+1}\,\Gamma_{a+2}\right) + (\zeta)^4\left(\gamma_{a+1}^1\, m_{a+2} - m_{a+1}\,\gamma_{a+2}\right) = 0
\end{aligned} \tag{A.4c}$$

$$\frac{d}{dt} m_c - (\Delta s)^{-1} \left( M_c^1 - M_c \right)$$
$$- \varepsilon_{abc} \left( p_a \gamma_b + \Gamma_a P_b^1 + \Omega_a M_b + m_a \omega_b^1 \right) = 0 \quad \text{(A.4d)}$$

In (A.4b) and (A.4c) above, index $a$ runs 1,2,3. Index $a+1$ and index $a+2$ take their values of modulo 3.

Assuming that $\zeta$ is pure imaginary, system (A.4) reduces to system (3.9).